\documentclass[12pt,preprint]{emulateapj}
\usepackage{epsfig}
\def\lap{\lower.5ex\hbox{$\; \buildrel < \over \sim \;$}}
\def\gap{\lower.5ex\hbox{$\; \buildrel > \over \sim \;$}}

\def\ergcm2s{${\rm erg\ cm^{-2}\ s^{-1}}$}

\def\ergscm2s{${\rm erg\ cm^{-2}\  s^{-1}}$}

\def\cm-2{${\rm cm^{-2}}$}

\begin{document}

\title{The History of Star Formation in Galaxy Disks in the Local
Volume as Measured by the ACS Nearby Galaxy Survey Treasury}

\author{Benjamin F. Williams\altaffilmark{1},
Julianne J. Dalcanton\altaffilmark{1},
L.~C. Johnson\altaffilmark{1},
Daniel R. Weisz\altaffilmark{1},
Anil C. Seth\altaffilmark{2,7},
Andrew Dolphin\altaffilmark{3},
Karoline M. Gilbert\altaffilmark{1,8},
Evan Skillman\altaffilmark{4},
Keith Rosema\altaffilmark{1},
Stephanie M. Gogarten\altaffilmark{1},
Jon Holtzman\altaffilmark{5},
Roelof S. de Jong\altaffilmark{6},
}

\altaffiltext{1}{Department of Astronomy, Box 351580, University of Washington, Seattle, WA 98195; ben@astro.washington.edu; jd@astro.washington.edu; lcj@astro.washington.edu; dweisz@astro.washington.edu; kgilbert@astro.washington.edu; krosema@astro.washington.edu; stephanie@astro.washington.edu}
\altaffiltext{2}{Harvard-Smithsonian Center for Astrophysics, 60 Garden Street, Cambridge, MA 02138; aseth@cfa.harvard.edu}
\altaffiltext{3}{Raytheon, 1151 E. Hermans Road, Tucson, AZ 85706; adolphin@raytheon.com}
\altaffiltext{4}{Department of Astronomy, University of Minnesota, 116 Church St. SE, Minneapolis, MN 55455; skillman@astro.umn.edu}
\altaffiltext{5}{Department of Astronomy, New Mexico State University, Box
30001, 1320 Frenger St., Las Cruces, NM 88003; holtz@nmsu.edu}
\altaffiltext{6}{Astrophysikalisches Institut Potsdam, Potsdam, Germany; rdejong@aip.de}
\altaffiltext{7}{CfA Fellow.}
\altaffiltext{8}{Hubble Fellow.}
  
\keywords{galaxies: stellar content ---  galaxies: evolution}

\begin{abstract}
 
We present a measurement of the age distribution of stars residing in spiral disks and dwarf galaxies.  We derive a complete star formation history of the $\sim$140~Mpc$^3$ covered by the volume-limited sample of galaxies in the Advanced Camera for Surveys (ACS) Nearby Galaxy Survey Treasury (ANGST). The total star formation rate density history ($\rho_{SFR}(t)$) is dominated by the large spirals in the volume, although the sample consists mainly of dwarf galaxies.  Our $\rho_{SFR}(t)$ shows a factor of $\sim$3 drop at $z\,\sim\,2$, in approximate agreement with results from other measurement techniques. While our results show that the overall $\rho_{SFR}(t)$ has decreased since $z\,\sim\,1$, the measured rates during this epoch are higher than those obtained from other measurement techniques.  This enhanced recent star formation rate appears to be largely due to an increase in the fraction of star formation contained in low-mass disks at recent times.  Finally, our results indicate that despite the differences at recent times, the epoch of formation of $\sim$50\% of the stellar mass in dwarf galaxies was similar to that of $\sim$50\% of the stellar mass in large spiral galaxies (z$\gap$2), despite the observed galaxy-to-galaxy diversity among the dwarfs.

\end{abstract}

\section{Introduction}

The star formation history of the universe ($\rho_{SFR}(t)$)
constrains models of structure formation, the assembly of galaxies,
metal production, and the epoch of reionization. Currently there are
testable models for the growth of structure in the universe on all
scales, for the flow of gas in and out of galaxies, and the conversion
of gas into stars.  If these models are correct, then we should expect
consistency between the models and observations of the rate at which
galaxies formed stars throughout cosmic time at all galaxy mass
scales.  Moreover, we should expect a similar level of consistency
between the observed {\it in situ} star formation rate (SFR) and the
stellar record and metal content of the universe at the present day.

With the combination of {\it HST}, large aperture redshifts surveys,
and well-calibrated photometric redshifts, there has been an explosion
of observational constraints on the SFR at high redshifts
\citep[e.g.][]{madau1996,connolly1997,lilly1996,steidel1999,fontana2003,iwata2003,bunker2004,giavalisco2004},
now pushing out to $z\gap8$ \citep{bouwens2010}.  These measurements
have been augmented by measurements of the obscured SFR due to
advances in the capabilities of long wavelength detectors allowing
measurements of high-$z$ star formation \citep{chapman2005}.  Such
assessment of the SFR has also been made at low redshifts by surveys
like the Sloan Digital Sky Survey \citep[SDSS][]{heavens2004} and the
Galaxy Evolution Explorer \citep[GALEX][]{schiminovich2005} all on the
basic properties of SFR versus time.

While early measurements of $\rho_{SFR}(t)$ showed a peak around
$z\sim1.5$
\citep[e.g.][]{madau1996,connolly1997,hopkins2001}, more
recent measurements have generally put the peak prior to $z\sim2$
\citep[e.g.][]{lanzetta2002,hopkins2004,chapman2005,bouwens2007,reddy2008},
including measurements combining the star formation histories (SFHs)
of Local Group galaxies.  Inside the LG \citet{hopkins2001}
found that $\rho_{SFR}(t)$ was broadly consistent with redshift
surveys with no significant contribution from dwarfs at any epoch, and
\citet{drozdovsky2008} found an excess of star formation in recent
epochs, dominated by the disk of the Milky Way, as well as a recent
increase in the contribution from dwarfs. \citet{weisz2011} found
little difference in the SFHs of the LG dwarfs and those in a larger
volume.  Only one recent measurement, based on integrated galaxy
spectra in the Sloan Digital Sky Survey \citep{heavens2004}, has found
a peak more recent than $z=2$.  Furthermore, recent analytical
calculations \citep{hernquist2003}, semi-analytic galaxy formation
models \citep{lacey2009}, and hydrodynamic simulations
\citep{springel2003} also generally put the peak earlier than
$z\sim2$.  For current WMAP cosmology, this places the peak at
lookback times $>$10 Gyr.  There is mounting evidence that low-mass
galaxies may have later formation times
\citep[i.e. ``down-sizing''][and many
others]{cowie1996,lilly2003,thomas2005,neistein2006},
which could in principle affect the location of the $\rho_{SFR}(t)$
peak. Unfortunately,direct constraints at high redshift are
challenging, given that all {\it in situ} redshift-based studies have
magnitude limits that prohibit the inclusion of low-mass galaxies in
their measurements.

Herein, we report our measurement $\rho_{SFR}(t)$ in our local volume
using resolved stellar populations from a volume-limited sample of
galaxy disks: the largest ever measured using resolved stellar
populations analysis.  Our approach has the benefit that we
simultaneously explore the past SFH and the present stellar record.
Moreover, we have complete, volume-limited sampling of the galaxy
population down to very low masses, rather than the high-mass galaxies
that dominate {\it in situ} high-redshift studies. The limitations of
the approach are difficulty resolving stars in massive spheroids, so
that we can only trace the SFH of disks and dwarf galaxies, and a
relatively short distance over which stars can be resolved with {\sl
HST}, so that we cannot measure a truly representative volume of the
universe.  Our results are generally consistent with those of recent
redshift surveys, and we conclude that with the currently-possible
depth of resolved stellar photometry over this volume, we cannot
resolve the age of the peak beyond placing it at $z\,>2$.

\section{Data Acquisition, Reduction, and Analysis}

All of the data for this study were analyzed through the ANGST
(GO-10915) and ACS Nearby Galaxies: Reduce, Reuse, Recycle (ANGRRR;
AR-10945) programs
\footnote{http://www.nearbygalaxies.org/}.  Our full sample is
detailed in Table~\ref{table} and in \citet[][]{dalcanton2009}.  The
motivation for the sample selection was to stay within a limited
volume.  We have included all galaxies inside of $\sim$3.5~Mpc, but
outside the LG \citep[as defined
by][$\sim$8~Mpc$^3$]{vandenbergh2000} and more than 20$^{\circ}$ from
the Galactic plane ($\sim$34\% of the sphere).  ANGST extended this
distance limit to $\sim$4~Mpc in the direction of the M81 group and
Sculptor filament to improve sampling of massive galaxies and dense
environments.  The Cen~A group, which is also within $\sim$4~Mpc, was
excluded due to its low galactic latitude and incomplete galaxy
census.  Our sample therefore covers only $\sim$70\% of the stellar
mass in the volume between 3.5 and 4.0 Mpc, making our effective
distance limit $\sim$3.8~Mpc.  Therefore, our net volume surveyed is
$\sim$140~Mpc$^3$.  This scale is smaller than the large scale
structure of the universe, and thus we may be sampling a
non-representative environment. Therefore differences between our
results and those of redshift surveys may be primarily due to such
sampling effects.

We exclude KDG73 and Sc22 as their revised distance moduli place them
beyond 4 Mpc, and we exclude BK6N and KKH57 due to poor data quality.
The program obtained ACS and WFPC2 imaging of a volume-limited sample
of galaxies. All photometry techniques are described in detail in
\citet{dalcanton2009} and K. Gilbert et al.(in preparation).  In
short, the photometry and artificial star tests were measured
simultaneously for all of the objects in the uncombined images using
the software packages HSTPHOT and DOLPHOT \citep{dolphin2000}, and the
output data were culled on signal-to-noise, sharpness, and crowding.

\subsection{SFH Determination}\label{uncertainty}

We measured the SFR and metallicity as a function of stellar age using
the software package MATCH \citep{dolphin2002}.  We fit the observed
CMDs (with magnitude cuts set to limits provided in Table~1) by
populating the stellar evolution models of \citet[][with updates in
\citealp{marigo2008,girardi2010}]{girardi2002} with a
\citet{salpeter1955} initial mass function (IMF).  We fixed the
distance and reddening to the \citet{dalcanton2009} values.  The best
fit provides the relative contribution of stars of each age and
metallicity in each field.

We then performed Monte Carlo fits by resampling the best-fitting
model 100 times. Then, when fitting the realizations, the systematic
errors are accounted for by introducing small random shifts in the
bolometric magnitudes and effective temperatures of the models.  These
shifts are introduced at the level of the differences between models
in the literature, and therefore serve as a proxy of the effects of
our choice of stellar evolution models.  From these tests we
calculated our uncertainties due to Poisson sampling, errors in
photometry, and systematic errors due to deficiencies in the stellar
evolution models as well as any offset in distance, reddening, and/or
zero-points.

Our time bins were chosen based on the features present in most of our
data.  In general, the main-sequence and blue He-burning sequences
provide high resolution time sensitivity for times more recent than
$\sim$400~Myr, After this epoch, our photometry contains only the RGB,
which contains degeneracies between age and metallicity
\citep{gallart2005}, and the AGB, which is generally poorly populated
and suffers from poorly-constrained models \citep{melbourne2010}, and
the red clump, which is dominated by old ($\gap$2 Gyr) stars. We
therefore limited ourselves to 4 bins in this large interval: one long
bin on each side of $\sim$1~Gyr ago (the epoch where our CMDs provide
the least information) in order to leverage more reliable age
information from before and after this period, and two more bins where
we have additional information from the red clump to help constrain
the age distribution (4--10 Gyr and 10--14 Gyr).

\begin{figure}
\centerline{\psfig{file=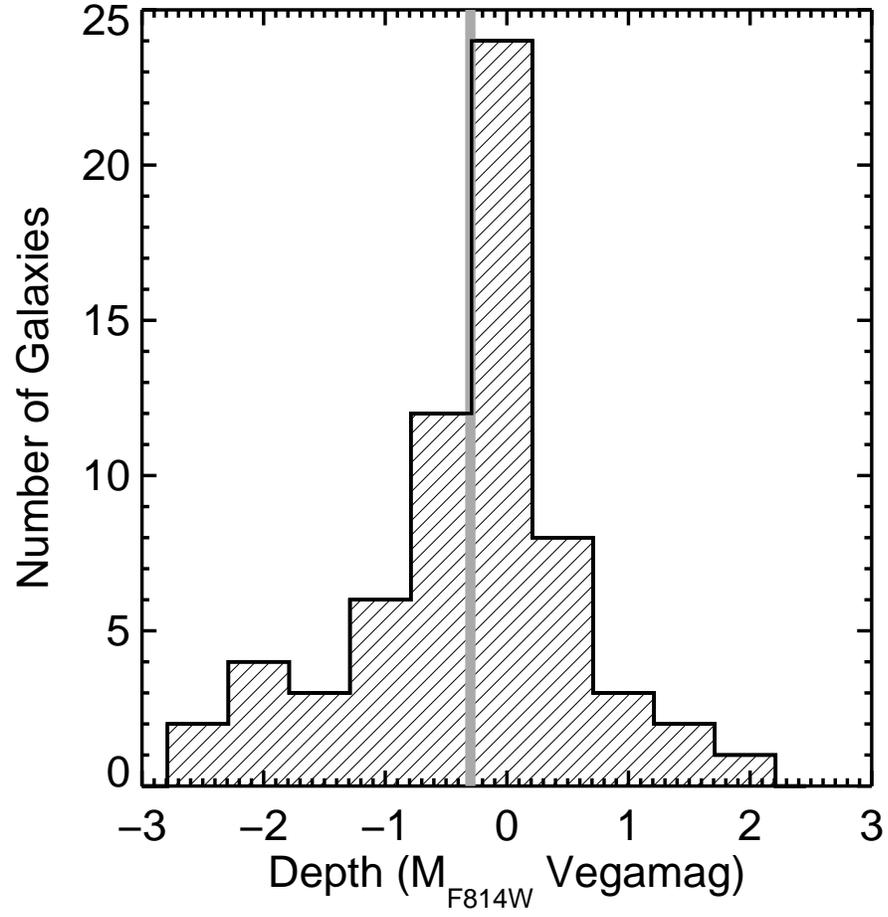,height=5.0in,angle=0}}
\caption{Histogram of the 50\% completeness limiting F814W absolute
magnitude for the sample galaxies.  These values have been corrected
for distance and extinction. The depth of 58\% is fainter than the red
clump in our M81 data (thick gray line).}
\label{hist}
\end{figure}

When determining the SFHs, we check the effects of our varying depth
and magnitude cuts for fitting each CMD.  Due to the failure of ACS
during the ANGST program, our photometry sample is heterogeneous and
of varying depth (see electronic Table~1).  Nevertheless, the bulk of
the stellar mass is in M81 and NGC253, which were both observed with
ACS, and every attempt was made to obtain a depth reaching the red
clump feature in the CMD with WFPC2 ($M_{F814W}\sim-0.3,
M_{F606W}\sim0.4$). Furthermore, our SFHs for ancient times ($>$2~Gyr)
were always taken from the deepest data available, which for most
low-mass galaxies, included information from the red clump feature
(see electronic Table~1 and Figure~\ref{hist}).  There is still no
guarantee that our data were all detailed enough to reliably separate
populations with ages slightly less than 10~Gyr from those with ages
slightly greater than 10~Gyr.

\subsection{SFH Scaling}

Our fields cover only a portion of each galaxy, making it necessary to
scale the measured SFH to the total galaxy mass.  Once we had measured
the SFHs for the deepest available field in each ANGST galaxy, we
renormalized the SFRs to a \citet{kroupa2002} IMF (divided by 2), and
we scaled them to the total stellar mass in each galaxy. We first
estimated the galaxy's total stellar mass from {\it Spitzer}
photometry (see below). The stellar mass contained in our ANGST field
was then calculated directly from our SFH.  The quotient of these
provided our estimate of the fraction of the galaxy's stellar mass
contained in our field
($SFR_{total}(t)=SFR_{field}(t)\times\frac{M_{star,3.6\mu}}{M_{star,SFH}}$).

To estimate each galaxy's total stellar mass, we used the total
Spitzer 3.6$\mu$m luminosity from the {\it Spitzer} Local Volume
Legacy Survey \citep{dale2009}.  To determine the appropriate
$M/L_{3.6\mu}$ to apply, we calculated $M/L_{3.6\mu}$ using stellar
masses determined by our CMD fits and Spitzer 3.6$\mu$ photometry of
the fields with good Spitzer coverage and minimal foreground
contamination.  In all cases with reliable Spitzer data covering our
field, $M/L_{3.6\mu}=$0.5$\pm$0.2. We therefore assumed
$M/L_{3.6\mu}$=0.5 to estimate each galaxy's total stellar mass.  We
verified that this assumption resulted in a total stellar masses of
M81 and NGC~253 less than the mass calculated from their rotation
curves \citep[as taken from][]{puche1991,adler1996}.  For large
spirals, the total stellar mass was further scaled by the galaxy's
disk/(bulge+disk) luminosity ratio from the literature to avoid
representing stellar mass in the bulge with the SFH of the disk.

For the largest ANGST galaxies, most of the young ($<$2~Gyr)
populations are not well-mixed and are not contained within a single
ANGST field.  Thus our assumption that the SFH in the deepest field
(which is located in the outer regions of the disk, where crowding is
minimized) is representative of the entire galaxy is not valid at
recent timescales that do not allow sufficient radial mixing. In these
cases, we therefore made use of shallower tilings that covered at
least half of the optical disk.  Where possible (M81, NGC~253, NGC~55,
and NGC~3109), we measured the recent SFHs ($<$2 Gyr; $z<$0.2) from
the shallower galaxy tilings.  These tilings cover the full extent of
the optical disk of M81 and half of the optical disk of the other 3
galaxies (stepping along the major axis from the center).  This method
provides a more realistic total recent SFH for these large disk
galaxies, requiring less scaling.  The case of M81 required no
scaling, and the other 3 galaxies required scaling only by a factor of
2 to account for the ANGST coverage.  However, since a significant
amount of star formation may be in unresolved clusters of stars and
extincted by dust, we still likely miss some fraction of the most
recent star formation.  We verified that the recent SFRs obtained by
this method were at least equal to the rates that would have come out
of scaling the SFH of the deep field by the total {\it Spitzer}
stellar mass of the galaxy in order to be sure that using the shallow
tiled data was, at the very least, not reducing the amount recent star
formation contained in these galaxies.

We note our measuring technique is not sensitive to the stellar
populations in bulges.  Because of severe crowding in the central
portions of the large galaxies, the photometry was too shallow to
produce reliable SFHs, and was therefore not included.  The galaxy
most strongly under-represented due to this bias against bulges is
M82, which is known to have a total current SFR
$\sim$1--6~M$_{\odot}$~yr$^{-1}$
\citep{young1996,bell2001a,kennicutt2003}, mainly in the central
500~pc of the galaxy.  On the other hand, the disk has a recent
($<$40~Myr) SFR in our measurement of just
0.2$\pm$0.1~M$_{\odot}$~yr$^{-1}$.  Our measured rate is similar
whether we scale the SFR as determined from our deepest data or sum
the recent SFRs from the available tilings.  The star formation is
mainly in the bulge, which dominates the galaxy luminosity
\citep[$L_{disk}/L_{bulge}\,\sim\,0.4$][]{ichikawa1995}.  For this
reason, $\rho_{SFR}$ for our most recent time bin is likely only a
lower limit.  In NGC~253, where the star formation is more widespread
throughout the disk, and the disk dominates the light
\citep[$L_{disk}/L_{bulge}\,\sim\,4$][]{simien1986}, our technique
recovered the recent SF better.  The current rate is
$\sim$0.7~M$_{\odot}$~yr$^{-1}$ \citep{young1996}, and we measure
0.4$\pm$0.05~M$_{\odot}$~yr$^{-1}$ (20--80~Myr). This value increases
to 0.8$\pm$0.1~M$_{\odot}$~yr$^{-1}$ when renormalized to a Salpeter
IMF.

When combining these final SFHs for the sample volume, we added
uncertainties in quadrature and increased the total uncertainty by a
factor of $\sqrt{2}$ to approximately account for additional
uncertainty from scaling and combining SFHs.  We also derived total
SFRs excluding M81 or NGC~253 to assess the sensitivity of our total
to these massive galaxies.  While this test removed a significant
fraction of the total stellar mass, the SFH remained essentially
unchanged in shape.  However, as shown in Figure~\ref{components}, M82
is disproportionately responsible for the high SFR $\sim$1--4~Gyr ago.

\section{Results}

\begin{figure}
\centerline{\psfig{file=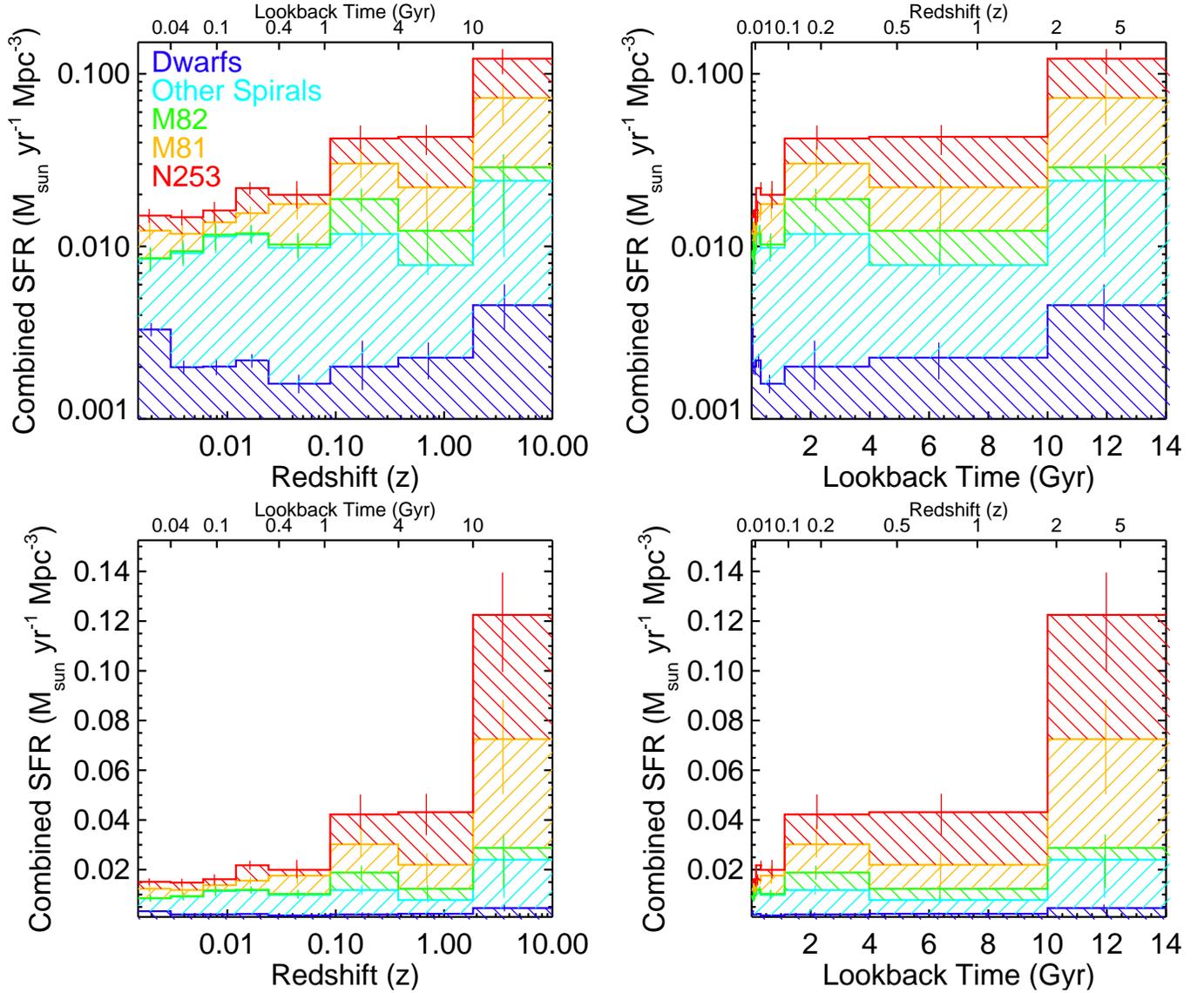,height=6.0in,angle=0}}
\caption{{\it Top Left:} The SFR density of the local volume.  Error
bars show the scaled root-sum-squared of the uncertainty estimates
from our measured SFHs (\S~\ref{uncertainty}).  Histograms denote
combined $\rho_{SFR}(t)$ including all 54 dwarf galaxies in the sample
(blue), 8 of the spiral disks (all but M82, M81, and NGC~253), M82
alone (green), M81 alone (yellow) and, completing the sample, NGC~253
alone (red).  {\it Other Panels:} Same as {\it Top Left} but with
linear lookback time (top right), linear SFR (bottom left), and both
(bottom right). We adopt a five-year WMAP \citep{dunkley2009}
cosmology for all conversions between time and redshift.  }
\label{components}
\end{figure}

\begin{figure}
\centerline{\psfig{file=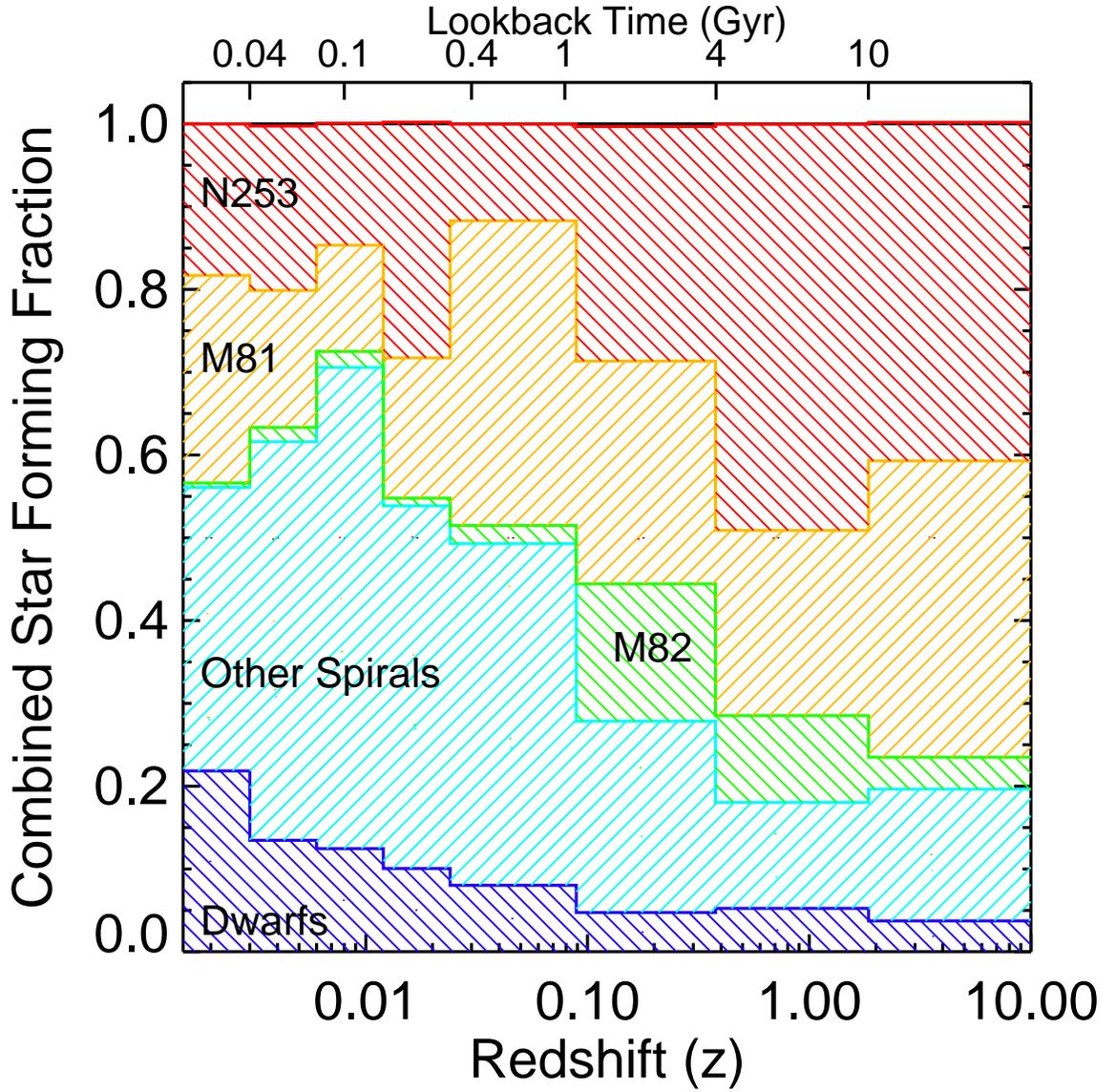,height=6.0in,angle=0}}
\caption{Fractional contribution of several components to the total
SFH of the survey volume. Colors are the same as in
Figure~\ref{components}.}
\label{frac}
\end{figure}

\begin{figure}
\centerline{\psfig{file=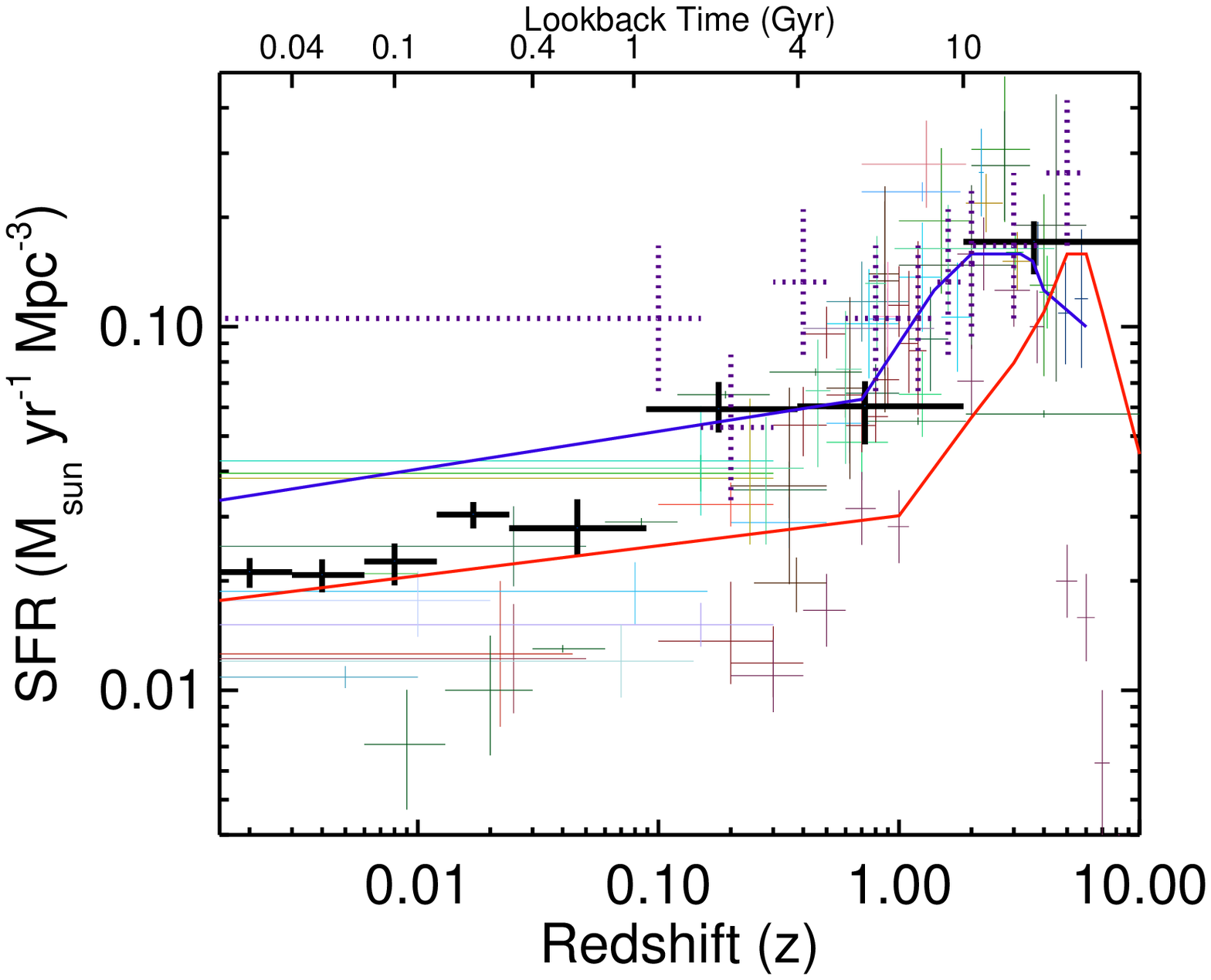,width=6.0in,angle=0}}
\caption{{\it Black Error Bars:} $\rho_{SFR}(t)$ measured from the
ANGST sample. Values have been scaled to reproduce the mean cosmic
stellar density for relevant comparison.  {\it Color Error Bars:}
Measurements taken from the compilation of \citet{hopkins2004} with
updates included in \citet{hopkins2007} and the results of
\citet{heavens2004}, \citet{reddy2008}, and \citet{bouwens2010}. {\it
Dotted Error Bars:} Measurements taken from the LG study of
\citet{drozdovsky2008}, scaled by a factor of 3 to compensate for the
overdensity of the LG.  {\it Lines:} Theoretical predictions
from semi-analytic galaxy evolution calculations
\citep[blue,][]{lacey2009} and hydrodynamic simulations
\citep[red,]{springel2003}.}
\label{literature}
\end{figure}

Our volume-limited $\rho_{SFR}(t)$, along with the contributions of
some major components (spirals, dwarfs, M81, and NGC253), is shown in
Figures~\ref{components} and \ref{frac}.  Overall, $\rho_{SFR}(t)$ in
the local volume is clearly dominated by that of the large spirals
NGC~253 and M81. In Figure~\ref{literature}, we plot $\rho_{SFR}(t)$,
scaled to match the cosmic mean stellar density, along with a
compilation of literature studies
\citep{hopkins2004,hopkins2007,reddy2008,heavens2004}, the LG
\citep{drozdovsky2008}, and two theoretical results
\citep{lacey2009,springel2003}.  The models are not easily compared to
our results because most of our time resolution comes at epochs more
recent than $z\,=\,1$ whereas the models typically provide one data
point per unit redshift.  However, overall it is encouraging that our
results are similar to the observational results already in the
literature, determined using other techniques, that our results are
similar to the SFH of the LG, and that our results fall
within the wide range of theoretical calculations.  We note that our
result differs from those of \citet{heavens2004}, which analyzed
spectra of $\sim$10$^5$ nearby galaxies and found the local
$\rho_{SFR}(t)$ peak at $z\,\sim\,0.6$.  While both measurements show
enhanced star formation at lookback times of $\sim$2--6 Gyr, the
discrepancy is in the $z\,>\,2$ ($>$10~Gyr) bin, where our measured
rate is a factor of 3 larger than theirs, placing our peak rate in
this oldest bin. Their sample and technique remain the only ones to
have measured such a low rate for $z\,>\,2$.

Despite the diversity of SFHs in local dwarf galaxies
\citep[$M_B>-18$][]{weisz2011a}, when one sums their total
$\rho_{SFR}(t)$, it is remarkably similar to that of the full galaxy
sample prior to $z\,\sim\,0.1$ ($>$1 Gyr lookback time).  For all but
this recent epoch, the $\rho_{SFR}(t)$ pattern for the dwarfs is
indistinguishable from that of the large spirals, which, in turn, is
indistinguishable from that of the sample volume.  Therefore, we do
not detect any significant difference between the formation times of
dwarf galaxies and large spirals, suggesting that any differences
occurred prior to $z\,\sim\,2$ ($>$10 Gyr ago).

At lookback times $\gap$4 Gyr ($z\,\gap\,0.5$), our total SFRs are
indistinguishable from those of most redshift surveys. However, our
recent SFRs are higher than those of redshift surveys, similar to
those of the LG \citep{drozdovsky2008}.  This result is
largly due to our inclusion of low-mass galaxies (other spirals and
dwarfs in Figures~\ref{components} and \ref{frac}), which contribute
50\% of the SFR at recent times.  Such low-mass galaxies are typically
not included in redshift surveys due to their low luminosities.
 
\section{Conclusions}

We have derived $\rho_{SFR}(t)$ for galaxy disks in our local volume
using resolved stellar photometry.  This measurement includes the
contribution by different galaxy types.  Our sample is dominated, in
number, by galaxies fainter than the limits of any available redshift
survey or high-$z$ {\sl HST} imaging.  Thus, our $\rho_{SFR}(t)$
represents a portion of parameter space not yet studied in detail
outside the LG.  However, our sample is lacking in massive
spheroids, whose mean $\rho_{SFR}(t)$ is more reliably determined from
the large samples included in redshift surveys and high-$z$ {\sl HST}
imaging.

We find that, while $\rho_{SFR}(t)$ is dominated by that of the
spirals, the overall shape of the measurement is robust against
removal of any single galaxy.  We also find that the combined
$\rho_{SFR}(t)$ of the dwarf galaxies is not significantly different
in shape from that of the larger spirals except for the most recent
Gyr, suggesting that $\sim$50\% of the stellar mass in both types of
systems formed prior to $z\sim2$.  This result is consistent with
those of \citet{thomas2005} for high-density environments.  

We have compared our measurements to those obtained by galaxy surveys
and analysis of the HDF and UDF.  We find overall agreement between
our results and those of galaxy surveys; however, our measurements do
not yet have the time resolution at epochs prior to 10~Gyr ($z\gap2$)
to resolve the peak in cosmic SFR density.  Our measurements suggest
that this peak lies prior to $z\,\sim\,2$, consistent with the most
recent {\it HST/WFC3} results, which place the peak at $z\,\sim\,4$
\citep{bouwens2010}. Finally, at recent times, the contribution of
low-mass galaxies to the total SFR has increased, resulting in a
higher total rate at recent times than observed by other methods.

Support for this work was provided by NASA through grants GO-10915,
GO-11986, and Hubble Fellowship grant (for KMG) HST-HF-51273.01
awarded by the Space Telescope Science Institute, which is operated by
the Association of Universities for Research in Astronomy, Inc., for
NASA, under contract NAS 5-26555.

\newpage

\LongTables

\begin{deluxetable}{lcccccccc}
\tablewidth{25cm}
\tablecaption{Data Used for SFH Determination}
\tabletypesize{\tiny}
\tablehead{
\colhead{Galaxy} &
\colhead{Proposal} &
\colhead{Target} &
\colhead{Camera} &
\colhead{Filter} &
\colhead{Exposure (s)}&
\colhead{Stars} &
\colhead{$m_{50\%}$} &
\colhead{$M_{50\%}$}
}
\startdata
Antlia/P29194 & 10210 & ANTLIA & ACS & F606W &    985 & 19226 &  28.01
&  2.18\\
Antlia/P29194 & 10210 & ANTLIA & ACS & F814W &   1174 & 19226 &  27.30 &  1.54\\
KK230 & 9771 & KK230 & ACS & F606W &   1200 & 4679 &  28.15 &  1.69\\
KK230 & 9771 & KK230 & ACS & F814W &    900 & 4679 &  27.08 &  0.64\\
E410-005/KK3 & 10503 & ESO410-005 & ACS & F606W &   8960 & 79952 &  28.85 &  2.39\\
E410-005/KK3 & 10503 & ESO410-005 & ACS & F814W &  22400 & 79952 &  27.92 &  1.48\\
E294-010 & 10503 & ESO294-010 & ACS & F606W &  13920 & 103465 &  28.95 &  2.52\\
E294-010 & 10503 & ESO294-010 & ACS & F814W &  27840 & 103465 &  28.51 &  2.08\\
GR8/DDO155 & 10915 & GR8 & ACS & F475W &   2244 & 22134 &  28.23 &  1.52\\
GR8/DDO155 & 10915 & GR8 & ACS & F814W &   2259 & 22134 &  27.46 &  0.80\\
N300 & 10915 & NGC0300-WIDE1 & ACS & F606W &   1515 & 224152 &  27.89 &  1.19\\
N300 & 10915 & NGC0300-WIDE1 & ACS & F814W &   1542 & 224152 &  27.00 &  0.31\\
DDO187 & 10210 & UGC9128 & ACS & F606W &    985 & 27599 &  27.88 &  1.02\\
DDO187 & 10210 & UGC9128 & ACS & F814W &   1174 & 27599 &  27.13 &  0.30\\
KKH98 & 10915 & KKH98 & ACS & F475W &   2265 & 10904 &  28.21 &  0.78\\
KKH98 & 10915 & KKH98 & ACS & F814W &   2280 & 10904 &  27.41 &  0.22\\
U8508 & 10915 & UGC8508 & ACS & F475W &   2280 & 45958 &  27.97 &  0.87\\
U8508 & 10915 & UGC8508 & ACS & F814W &   2349 & 45958 &  27.32 &  0.25\\
DDO190/U9240 & 10915 & DDO190 & ACS & F606W &   2301 & 105910 &  28.15 &  0.89\\
DDO190/U9240 & 10915 & DDO190 & ACS & F814W &   2265 & 105910 &  27.35 &  0.10\\
DDO113/KDG90 & 10915 & DDO113 & ACS & F475W &   2265 & 21150 &  28.21 &  0.85\\
DDO113/KDG90 & 10915 & DDO113 & ACS & F814W &   2280 & 21150 &  27.43 &  0.11\\
DDO181/U8651 & 10210 & UGC8651 & ACS & F606W &   1016 & 41855 &  27.96 &  0.54\\
DDO181/U8651 & 10210 & UGC8651 & ACS & F814W &   1209 & 41855 &  27.12 & -0.29\\
N3741 & 10915 & NGC3741 & ACS & F475W &   2262 & 29476 &  28.04 &  0.54\\
N3741 & 10915 & NGC3741 & ACS & F814W &   2331 & 29476 &  27.33 & -0.12\\
N4163 & 10915 & NGC4163 & ACS & F606W &   2292 & 97632 &  28.05 &  0.63\\
N4163 & 10915 & NGC4163 & ACS & F814W &   2250 & 97632 &  27.32 & -0.08\\
UA292 & 10915 & UGCA292 & ACS & F606W &    926 & 8913 &  27.89 &  0.39\\
UA292 & 10915 & UGCA292 & ACS & F814W &   2274 & 8913 &  27.39 & -0.10\\
U8833 & 10210 & UGC8833 & ACS & F606W &    998 & 19438 &  27.89 &  0.39\\
U8833 & 10210 & UGC8833 & ACS & F814W &   1189 & 19438 &  27.11 & -0.38\\
DDO183/U8760 & 10210 & UGC8760 & ACS & F606W &    998 & 36852 &  27.86 &  0.30\\
DDO183/U8760 & 10210 & UGC8760 & ACS & F814W &   1189 & 36852 &  27.08 & -0.46\\
N2366 & 10605 & NGC-2366-2 & ACS & F555W &   4780 & 237638 &  28.06 &  0.41\\
N2366 & 10605 & NGC-2366-2 & ACS & F814W &   4780 & 237638 &  27.53 & -0.06\\
DDO44/KK61 & 10915 & DDO44 & ACS & F475W &   2361 & 34481 &  28.31 &  0.62\\
DDO44/KK61 & 10915 & DDO44 & ACS & F814W &   2430 & 34481 &  27.55 & -0.06\\
DDO6 & 10915 & DDO6 & ACS & F475W &   2250 & 23799 &  28.34 &  0.66\\
DDO6 & 10915 & DDO6 & ACS & F814W &   2268 & 23799 &  27.53 & -0.12\\
KKH37/Mai16 & 10915 & KKH37 & ACS & F475W &   2469 & 15361 &  28.31 &  0.36\\
KKH37/Mai16 & 10915 & KKH37 & ACS & F814W &   2541 & 15361 &  27.58 & -0.22\\
HoII/DDO50 & 10605 & UGC-4305-1 & ACS & F555W &   4660 & 248011 &  27.95 &  0.19\\
HoII/DDO50 & 10605 & UGC-4305-1 & ACS & F814W &   4660 & 248011 &  27.38 & -0.33\\
KDG2/E540-030 & 10503 & ESO540-030 & ACS & F606W &   6720 & 16964 &  28.67 &  0.94\\
KDG2/E540-030 & 10503 & ESO540-030 & ACS & F814W &   6720 & 16964 &  27.81 &  0.11\\
E540-032/FG24 & 10503 & ESO540-032 & ACS & F606W &   8960 & 34278 &  28.87 &  1.14\\
E540-032/FG24 & 10503 & ESO540-032 & ACS & F814W &   4480 & 34278 &  27.72 &  0.01\\
FM1 & 9884 & M81F6D1 & ACS & F606W &  17200 & 19373 &  28.86 &  0.98\\
FM1 & 9884 & M81F6D1 & ACS & F814W &   9000 & 19373 &  27.79 & -0.02\\
KK77 & 9884 & M81F12D1 & ACS & F606W &  17200 & 59039 &  29.06 &  0.93\\
KK77 & 9884 & M81F12D1 & ACS & F814W &   9000 & 59039 &  27.98 & -0.01\\
KDG63/KK83 & 9884 & DDO71 & ACS & F606W &  17200 & 57133 &  28.92 &  0.92\\
KDG63/KK83 & 9884 & DDO71 & ACS & F814W &   9000 & 57133 &  28.01 &  0.10\\
M82 & 10776 & M82-POS4 & ACS & F555W &   1360 & 31842 &  27.29 & -0.51\\
M82 & 10776 & M82-POS4 & ACS & F814W &    700 & 31842 &  26.35 & -1.43\\
KDG52 & 10605 & MESSIER-081-DWARF-A & ACS & F555W &   5914 & 20335 &  28.54 &  0.72\\
KDG52 & 10605 & MESSIER-081-DWARF-A & ACS & F814W &   5936 & 20335 &  27.99 &  0.20\\
DDO53 & 10605 & UGC-04459 & ACS & F555W &   4768 & 80038 &  28.32 &  0.44\\
DDO53 & 10605 & UGC-04459 & ACS & F814W &   4768 & 80038 &  27.76 & -0.07\\
N2976 & 10915 & NGC2976-DEEP & ACS & F606W &  18716 & 105537 &  29.21 &  1.25\\
N2976 & 10915 & NGC2976-DEEP & ACS & F814W &  27091 & 105537 &  28.59 &  0.70\\
KDG61/KK81 & 9884 & M81K61 & ACS & F606W &  17200 & 80746 &  29.06 &  1.07\\
KDG61/KK81 & 9884 & M81K61 & ACS & F814W &   9000 & 80746 &  28.10 &  0.18\\
M81 & 10915 & M81-DEEP & ACS & F606W &  24132 & 171101 &  29.69 &  1.65\\
M81 & 10915 & M81-DEEP & ACS & F814W &  29853 & 171101 &  29.04 &  1.08\\
M81 & 10584 & M81-FIELD-10 & ACS & F435W &   1200 & 23950 &  27.49 & -0.68\\
M81 & 10584 & M81-FIELD-10 & ACS & F606W &   1200 & 23950 &  27.35 & -0.69\\
M81 & 10584 & M81-FIELD-11 & ACS & F435W &   1200 & 64029 &  27.43 & -0.74\\
M81 & 10584 & M81-FIELD-11 & ACS & F606W &   1200 & 64029 &  27.18 & -0.86\\
M81 & 10584 & M81-FIELD-12 & ACS & F435W &   1200 & 94882 &  27.17 & -1.00\\
M81 & 10584 & M81-FIELD-12 & ACS & F606W &   1200 & 94882 &  26.63 & -1.41\\
M81 & 10584 & M81-FIELD-13 & ACS & F435W &   1200 & 100833 &  27.21 & -0.96\\
M81 & 10584 & M81-FIELD-13 & ACS & F606W &   1200 & 100833 &  26.75 & -1.29\\
M81 & 10584 & M81-FIELD-14 & ACS & F435W &   1200 & 45241 &  27.51 & -0.66\\
M81 & 10584 & M81-FIELD-14 & ACS & F606W &   1200 & 45241 &  27.31 & -0.73\\
M81 & 10584 & M81-FIELD-15 & ACS & F435W &   1200 & 48611 &  27.38 & -0.79\\
M81 & 10584 & M81-FIELD-15 & ACS & F606W &   1200 & 48611 &  27.21 & -0.83\\
M81 & 10584 & M81-FIELD-16 & ACS & F435W &   1200 & 117652 &  26.93 & -1.24\\
M81 & 10584 & M81-FIELD-16 & ACS & F606W &   1200 & 117652 &  26.40 & -1.64\\
M81 & 10584 & M81-FIELD-17 & ACS & F435W &   1200 & 121280 &  27.12 & -1.05\\
M81 & 10584 & M81-FIELD-17 & ACS & F606W &   1200 & 121280 &  26.43 & -1.61\\
M81 & 10584 & M81-FIELD-18 & ACS & F435W &   1200 & 71571 &  27.39 & -0.78\\
M81 & 10584 & M81-FIELD-18 & ACS & F606W &   1200 & 71571 &  27.19 & -0.85\\
M81 & 10584 & M81-FIELD-19 & ACS & F435W &   1200 & 46894 &  27.46 & -0.71\\
M81 & 10584 & M81-FIELD-19 & ACS & F606W &   1200 & 46894 &  27.12 & -0.92\\
M81 & 10584 & M81-FIELD-20 & ACS & F435W &   1200 & 73703 &  27.28 & -0.89\\
M81 & 10584 & M81-FIELD-20 & ACS & F606W &   1200 & 73703 &  26.91 & -1.13\\
M81 & 10584 & M81-FIELD-21 & ACS & F435W &   1200 & 68639 &  27.26 & -0.91\\
M81 & 10584 & M81-FIELD-21 & ACS & F606W &   1200 & 68639 &  26.88 & -1.16\\
M81 & 10584 & M81-FIELD-22 & ACS & F435W &   1200 & 28694 &  27.56 & -0.61\\
M81 & 10584 & M81-FIELD-22 & ACS & F606W &   1200 & 28694 &  27.39 & -0.65\\
M81 & 10584 & M81-FIELD-23 & ACS & F435W &   1200 & 27768 &  27.61 & -0.56\\
M81 & 10584 & M81-FIELD-23 & ACS & F606W &   1200 & 27768 &  27.46 & -0.58\\
M81 & 10584 & M81-FIELD-24 & ACS & F435W &   1200 & 65533 &  27.48 & -0.69\\
M81 & 10584 & M81-FIELD-24 & ACS & F606W &   1200 & 65533 &  27.30 & -0.74\\
M81 & 10584 & M81-FIELD-25 & ACS & F435W &   1200 & 66310 &  27.43 & -0.74\\
M81 & 10584 & M81-FIELD-25 & ACS & F606W &   1200 & 66310 &  27.21 & -0.83\\
M81 & 10584 & M81-FIELD-26 & ACS & F435W &   1200 & 13063 &  27.65 & -0.52\\
M81 & 10584 & M81-FIELD-26 & ACS & F606W &   1200 & 13063 &  27.36 & -0.68\\
M81 & 10584 & M81-FIELD-27 & ACS & F606W &   1580 & 136043 &  27.99 & -0.05\\
M81 & 10584 & M81-FIELD-27 & ACS & F814W &   1595 & 136043 &  27.20 & -0.76\\
M81 & 10584 & M81-FIELD-28 & ACS & F606W &   1580 & 190198 &  27.96 & -0.08\\
M81 & 10584 & M81-FIELD-28 & ACS & F814W &   1595 & 190198 &  27.18 & -0.78\\
M81 & 10584 & M81-FIELD-29 & ACS & F606W &   1580 & 80855 &  28.05 &  0.01\\
M81 & 10584 & M81-FIELD-29 & ACS & F814W &   1595 & 80855 &  27.25 & -0.71\\
M81 & 10584 & M81-FIELD-1 & ACS & F606W &   1580 & 84484 &  28.01 & -0.03\\
M81 & 10584 & M81-FIELD-1 & ACS & F814W &   1595 & 84484 &  27.16 & -0.80\\
M81 & 10584 & M81-FIELD-2 & ACS & F606W &   1580 & 119880 &  28.03 & -0.01\\
M81 & 10584 & M81-FIELD-2 & ACS & F814W &   1595 & 119880 &  27.17 & -0.79\\
M81 & 10584 & M81-FIELD-3 & ACS & F435W &   1200 & 12807 &  27.59 & -0.58\\
M81 & 10584 & M81-FIELD-3 & ACS & F606W &   1200 & 12807 &  27.53 & -0.51\\
M81 & 10584 & M81-FIELD-4 & ACS & F435W &   1200 & 42770 &  27.48 & -0.69\\
M81 & 10584 & M81-FIELD-4 & ACS & F606W &   1200 & 42770 &  27.39 & -0.65\\
M81 & 10584 & M81-FIELD-5 & ACS & F435W &   1200 & 57981 &  27.49 & -0.68\\
M81 & 10584 & M81-FIELD-5 & ACS & F606W &   1200 & 57981 &  27.37 & -0.67\\
M81 & 10584 & M81-FIELD-6 & ACS & F435W &   1200 & 19856 &  27.61 & -0.56\\
M81 & 10584 & M81-FIELD-6 & ACS & F606W &   1200 & 19856 &  27.33 & -0.71\\
M81 & 10584 & M81-FIELD-7 & ACS & F435W &   1200 & 25584 &  27.51 & -0.66\\
M81 & 10584 & M81-FIELD-7 & ACS & F606W &   1200 & 25584 &  27.34 & -0.70\\
M81 & 10584 & M81-FIELD-8 & ACS & F435W &   1200 & 120790 &  27.37 & -0.80\\
M81 & 10584 & M81-FIELD-8 & ACS & F606W &   1200 & 120790 &  26.99 & -1.05\\
M81 & 10584 & M81-FIELD-9 & ACS & F435W &   1200 & 103540 &  27.32 & -0.85\\
M81 & 10584 & M81-FIELD-9 & ACS & F606W &   1200 & 103540 &  27.05 & -0.99\\
N247 & 10915 & NGC0247-WIDE1 & ACS & F606W &   2280 & 193431 &  28.01 &  0.14\\
N247 & 10915 & NGC0247-WIDE1 & ACS & F814W &   2250 & 193431 &  27.23 & -0.62\\
HoIX/DDO66 & 10605 & UGC-5336 & ACS & F555W &   4768 & 57610 &  28.44 &  0.32\\
HoIX/DDO66 & 10605 & UGC-5336 & ACS & F814W &   4768 & 57610 &  27.91 & -0.09\\
KDG64/KK85 & 9884 & M81K64 & ACS & F606W &  17200 & 68420 &  29.19 &  1.18\\
KDG64/KK85 & 9884 & M81K64 & ACS & F814W &   9000 & 68420 &  28.35 &  0.40\\
IKN & 9771 & IKN & ACS & F606W &   1200 & 24626 &  28.03 & -0.22\\
IKN & 9771 & IKN & ACS & F814W &    900 & 24626 &  26.95 & -1.16\\
DDO78/KK89 & 10915 & DDO78 & ACS & F475W &   2274 & 36430 &  28.19 &  0.23\\
DDO78/KK89 & 10915 & DDO78 & ACS & F814W &   2292 & 36430 &  27.54 & -0.37\\
N3077 & 10915 & NGC3077-WIDE1 & ACS & F606W &   1596 & 442068 &  26.84 & -1.27\\
N3077 & 10915 & NGC3077-WIDE1 & ACS & F814W &   1622 & 442068 &  26.34 & -1.70\\
HoI/DDO63 & 10605 & UGC-5139 & ACS & F555W &   4446 & 123920 &  28.34 &  0.25\\
HoI/DDO63 & 10605 & UGC-5139 & ACS & F814W &   5936 & 123920 &  27.88 & -0.14\\
A0952+69 & 10915 & A0952 & ACS & F475W &   2250 & 7810 &  28.42 &  0.15\\
A0952+69 & 10915 & A0952 & ACS & F814W &   2265 & 7810 &  27.63 & -0.48\\
N253 & 10523 & NGC0253-HALO-2 & ACS & F606W &    680 & 11830 &  27.56 & -0.47\\
N253 & 10523 & NGC0253-HALO-2 & ACS & F814W &    680 & 11830 &  26.66 & -1.35\\
N253 & 10523 & NGC0253-HALO-11 & ACS & F606W &    680 & 27423 &  27.47 & -0.56\\
N253 & 10523 & NGC0253-HALO-11 & ACS & F814W &    680 & 27423 &  26.57 & -1.44\\
N253 & 10523 & NGC0253-HALO-12 & ACS & F606W &    680 & 8326 &  27.70 & -0.33\\
N253 & 10523 & NGC0253-HALO-12 & ACS & F814W &    680 & 8326 &  26.73 & -1.28\\
N253 & 10523 & NGC0253-HALO-17 & ACS & F606W &    680 & 62499 &  27.62 & -0.41\\
N253 & 10523 & NGC0253-HALO-17 & ACS & F814W &    680 & 62499 &  26.68 & -1.33\\
N253 & 10523 & NGC0253-HALO-18 & ACS & F606W &    680 & 42574 &  27.62 & -0.41\\
N253 & 10523 & NGC0253-HALO-18 & ACS & F814W &    680 & 42574 &  26.62 & -1.39\\
N253 & 10915 & NGC0253-WIDE1 & ACS & F606W &   2283 & 293299 &  28.04 &  0.01\\
N253 & 10915 & NGC0253-WIDE1 & ACS & F814W &   2253 & 293299 &  27.26 & -0.75\\
N253 & 10915 & NGC0253-WIDE2 & ACS & F606W &   1508 & 435333 &  27.21 & -0.82\\
N253 & 10915 & NGC0253-WIDE2 & ACS & F814W &   1534 & 435333 &  26.50 & -1.51\\
N253 & 10915 & NGC0253-WIDE3 & ACS & F606W &   1508 & 427307 &  26.52 & -1.51\\
N253 & 10915 & NGC0253-WIDE3 & ACS & F814W &   1534 & 427307 &  25.79 & -2.22\\
N253 & 10915 & NGC0253-WIDE4 & ACS & F606W &   1508 & 417964 &  26.13 & -1.90\\
N253 & 10915 & NGC0253-WIDE4 & ACS & F814W &   1534 & 417964 &  25.32 & -2.69\\
N253 & 10915 & NGC0253-WIDE5 & ACS & F606W &   1508 & 348456 &  25.66 & -2.37\\
N253 & 10915 & NGC0253-WIDE5 & ACS & F814W &   1534 & 348456 &  24.46 & -3.55\\
HS117 & 9771 & HS117 & ACS & F606W &   1200 & 7317 &  28.01 & -0.31\\
HS117 & 9771 & HS117 & ACS & F814W &    900 & 7317 &  27.05 & -1.16\\
DDO82 & 10915 & DDO82 & ACS & F606W &   2454 & 172885 &  28.13 & -0.01\\
DDO82 & 10915 & DDO82 & ACS & F814W &   2442 & 172885 &  27.46 & -0.64\\
BK3N & 10915 & BK3N & ACS & F475W &   2250 & 8164 &  28.40 &  0.06\\
BK3N & 10915 & BK3N & ACS & F814W &   2265 & 8164 &  27.58 & -0.60\\
I2574 & 9755 & IC2574-SGS & ACS & F555W &   6400 & 342454 &  28.14 & -0.01\\
I2574 & 9755 & IC2574-SGS & ACS & F814W &   6400 & 342454 &  27.62 & -0.47\\
SexA/DDO75 & 7496 & DDO75 & WFPC2 & F555W &  19200 & 33295 &  27.37 &  1.62\\
SexA/DDO75 & 7496 & DDO75 & WFPC2 & F814W &  38400 & 33295 &  26.63 &  0.94\\
N3109 & 10915 & NGC3109-DEEP & WFPC2  & F606W &   2700 & 13262 &  26.75 &  0.93\\
N3109 & 10915 & NGC3109-DEEP & WFPC2  & F814W &   3900 & 13262 &  26.00 &  0.25\\
N3109 & 11307 & NGC3109-WIDE1 & WFPC2 & F606W &   2700 & 21710 &  26.84 &  1.02\\
N3109 & 11307 & NGC3109-WIDE1 & WFPC2 & F814W &   3900 & 21710 &  25.85 &  0.10\\
N3109 & 11307 & NGC3109-WIDE2 & WFPC2 & F606W &   2700 & 30297 &  26.73 &  0.91\\
N3109 & 11307 & NGC3109-WIDE2 & WFPC2 & F814W &   3900 & 30297 &  25.76 &  0.01\\
N3109 & 11307 & NGC3109-WIDE3 & WFPC2 & F606W &   2400 & 34166 &  26.52 &  0.70\\
N3109 & 11307 & NGC3109-WIDE3 & WFPC2 & F814W &   2400 & 34166 &  25.47 & -0.28\\
N3109 & 11307 & NGC3109-WIDE4 & WFPC2 & F606W &   2400 & 42509 &  26.11 &  0.29\\
N3109 & 11307 & NGC3109-WIDE4 & WFPC2 & F814W &   2400 & 42509 &  25.26 & -0.49\\
SexB/DDO70 & 10915 & SEXB-DEEP & WFPC2 & F606W &   2700 & 29624 &  26.75 &  0.99\\
SexB/DDO70 & 10915 & SEXB-DEEP & WFPC2 & F814W &   3900 & 29624 &  25.92 &  0.19\\
KKR25 & 8601 & KKR25 & WFPC2 & F606W &    600 & 923 &  26.13 & -0.24\\
KKR25 & 8601 & KKR25 & WFPC2 & F814W &    600 & 923 &  24.98 & -1.39\\
I5152/E237-27 & 10915 & IC5152-DEEP & WFPC2 & F606W &   4800 & 325 &  27.32 &  0.67\\
I5152/E237-27 & 10915 & IC5152-DEEP & WFPC2 & F814W &   9600 & 325 &  26.45 & -0.18\\
N55 & 11307 & NGC0055-WIDE1 & WFPC2 & F606W &   2000 & 37013 &  26.61 & -0.06\\
N55 & 11307 & NGC0055-WIDE1 & WFPC2 & F814W &   3700 & 37013 &  25.74 & -0.92\\
N55 & 11307 & NGC0055-WIDE2 & WFPC2 & F606W &   1800 & 36771 &  26.41 & -0.26\\
N55 & 11307 & NGC0055-WIDE2 & WFPC2 & F814W &   2600 & 36771 &  25.56 & -1.10\\
N55 & 11307 & NGC0055-WIDE3 & WFPC2 & F606W &   2700 & 47212 &  26.01 & -0.66\\
N55 & 11307 & NGC0055-WIDE3 & WFPC2 & F814W &   3900 & 47212 &  25.25 & -1.41\\
N55 & 11307 & NGC0055-WIDE4 & WFPC2 & F606W &   2700 & 55453 &  25.58 & -1.09\\
N55 & 11307 & NGC0055-WIDE4 & WFPC2 & F814W &   3900 & 55453 &  24.84 & -1.82\\
N55 & 11307 & NGC0055-WIDE5 & WFPC2 & F606W &   2700 & 54136 &  25.55 & -1.12\\
N55 & 11307 & NGC0055-WIDE5 & WFPC2 & F814W &   3900 & 54136 &  24.73 & -1.93\\
N55 & 10915 & NGC0055-DEEP & WFPC2 & F606W &   6000 & 17055 &  27.50 &  0.83\\
N55 & 10915 & NGC0055-DEEP & WFPC2 & F814W &  10800 & 17055 &  26.69 &  0.03\\
UA438 & 8192 & E407-G18 & WFPC2 & F606W &    600 & 5016 &  26.04 & -0.74\\
UA438 & 8192 & E407-G18 & WFPC2 & F814W &    600 & 5016 &  24.97 & -1.80\\
DDO125/U7577 & 8601 & UGC7577 & WFPC2 & F606W &    600 & 11520 &  26.02 & -1.07\\
DDO125/U7577 & 8601 & UGC7577 & WFPC2 & F814W &    600 & 11520 &  24.91 & -2.15\\
KKH86 & 8601 & KKH71 & WFPC2 & F606W &    600 & 727 &  26.19 & -0.97\\
KKH86 & 8601 & KKH71 & WFPC2 & F814W &    600 & 727 &  25.08 & -2.06\\
DDO99/U6817 & 8601 & UGC6817 & WFPC2 & F606W &    600 & 6536 &  26.13 & -1.05\\
DDO99/U6817 & 8601 & UGC6817 & WFPC2 & F814W &    600 & 6536 &  24.98 & -2.18\\
N4214 & 10915 & NGC4214-DEEP & WFPC2 & F606W &  15600 & 16806 &  27.97 &  0.57\\
N4214 & 10915 & NGC4214-DEEP & WFPC2 & F814W &  31200 & 16806 &  27.20 & -0.18\\
N404 & 10915 & NGC0404-DEEP & WFPC2 & F606W &  39000 & 40793 &  27.25 & -0.35\\
N404 & 10915 & NGC0404-DEEP & WFPC2 & F814W &  75400 & 40793 &  26.76 & -0.78\\
E321-014 & 8601 & PGC39032 & WFPC2 & F606W &    600 & 1745 &  25.98 & -1.81\\
E321-014 & 8601 & PGC39032 & WFPC2 & F814W &    600 & 1745 &  24.91 & -2.79\\
U4483 & 8769 & UGC4483 & WFPC2 & F555W &   9500 & 6634 &  27.67 &  0.02\\
U4483 & 8769 & UGC4483 & WFPC2 & F814W &   6900 & 6634 &  26.37 & -1.23\\
N4190 & 10905 & NGC-4190 & WFPC2 & F606W &   2200 & 12549 &  26.74 & -1.07\\
N4190 & 10905 & NGC-4190 & WFPC2 & F814W &   2200 & 12549 &  25.46 & -2.32\\
F8D1 & 5898 & GAL-094447+672619 & WFPC2 & F555W &   9000 & 14226 &  27.84 & -0.37\\
F8D1 & 5898 & GAL-094447+672619 & WFPC2 & F814W &  15200 & 14226 &  27.05 & -1.02\\
BK5N & 6964 & GAL-100441+681522 & WFPC2 & F555W &  15600 & 2332 &  27.78 & -0.31\\
BK5N & 6964 & GAL-100441+681522 & WFPC2 & F814W &  21340 & 2332 &  26.95 & -1.05\\
N2403 & 10915 & NGC2403-DEEP & WFPC2 & F606W &  32400 & 30617 &  28.05 &  0.34\\
N2403 & 10915 & NGC2403-DEEP & WFPC2 & F814W &  62100 & 30617 &  27.20 & -0.47\\
MCG9-20-131 & 11986 & MCG9-20-131 & WFPC2 & F606W &  10700 & 9088 &  27.89 & 0.01\\
MCG9-20-131 & 11986 & MCG9-20-131 & WFPC2 & F814W &  19300 & 9088 &  26.82 & -1.12\\
\enddata
\label{table}
\end{deluxetable}

\begin{thebibliography}{}

\bibitem[\protect\citeauthoryear{{Adler} \& {Westpfahl}}{{Adler} \&
  {Westpfahl}}{1996}]{adler1996}
{Adler}, D.~S.,  \& {Westpfahl}, D.~J. 1996, \aj, 111, 735

\bibitem[\protect\citeauthoryear{{Bell} \& {Kennicutt}}{{Bell} \&
  {Kennicutt}}{2001}]{bell2001a}
{Bell}, E.~F.,  \& {Kennicutt}, R.~C., Jr. 2001, \apj, 548, 681

\bibitem[\protect\citeauthoryear{{Bouwens} et~al.}{{Bouwens}
  et~al.}{2007}]{bouwens2007}
{Bouwens}, R.~J., {Illingworth}, G.~D., {Franx}, M.,  \& {Ford}, H. 2007, \apj,
  670, 928

\bibitem[\protect\citeauthoryear{{Bouwens} et~al.}{{Bouwens}
  et~al.}{2010}]{bouwens2010}
{Bouwens}, R.~J., et~al. 2010, \apjl, 709, L133

\bibitem[\protect\citeauthoryear{{Bunker} et~al.}{{Bunker}
  et~al.}{2004}]{bunker2004}
{Bunker}, A.~J., {Stanway}, E.~R., {Ellis}, R.~S.,  \& {McMahon}, R.~G. 2004,
  \mnras, 355, 374

\bibitem[\protect\citeauthoryear{{Chapman} et~al.}{{Chapman}
  et~al.}{2005}]{chapman2005}
{Chapman}, S.~C., {Blain}, A.~W., {Smail}, I.,  \& {Ivison}, R.~J. 2005, \apj,
  622, 772

\bibitem[\protect\citeauthoryear{{Connolly} et~al.}{{Connolly}
  et~al.}{1997}]{connolly1997}
{Connolly}, A.~J., {Szalay}, A.~S., {Dickinson}, M., {Subbarao}, M.~U.,  \&
  {Brunner}, R.~J. 1997, \apjl, 486, L11

\bibitem[\protect\citeauthoryear{{Cowie} et~al.}{{Cowie}
  et~al.}{1996}]{cowie1996}
{Cowie}, L.~L., {Songaila}, A., {Hu}, E.~M.,  \& {Cohen}, J.~G. 1996, \aj, 112,
  839

\bibitem[\protect\citeauthoryear{{Dalcanton} et~al.}{{Dalcanton}
  et~al.}{2009}]{dalcanton2009}
{Dalcanton}, J.~J., et~al. 2009, \apjs, 183, 67

\bibitem[\protect\citeauthoryear{{Dale} et~al.}{{Dale} et~al.}{2009}]{dale2009}
{Dale}, D.~A., et~al. 2009, \apj, 703, 517

\bibitem[\protect\citeauthoryear{{Dolphin}}{{Dolphin}}{2000}]{dolphin2000}
{Dolphin}, A.~E. 2000, \pasp, 112, 1383

\bibitem[\protect\citeauthoryear{{Dolphin}}{{Dolphin}}{2002}]{dolphin2002}
{Dolphin}, A.~E. 2002, \mnras, 332, 91

\bibitem[\protect\citeauthoryear{{Drozdovsky} et~al.}{{Drozdovsky}
  et~al.}{2008}]{drozdovsky2008}
{Drozdovsky}, I., {Hopkins}, A., {Aparicio}, A.,  \& {Gallart}, C. 2008,
  {Matching the Local and Cosmic Star Formation Histories}, ed. {Koribalski,
  B.~S.~\& Jerjen, H.} (Springer Netherlands), 143

\bibitem[\protect\citeauthoryear{{Dunkley} et~al.}{{Dunkley}
  et~al.}{2009}]{dunkley2009}
{Dunkley}, J., et~al. 2009, \apjs, 180, 306

\bibitem[\protect\citeauthoryear{{Fontana} et~al.}{{Fontana}
  et~al.}{2003}]{fontana2003}
{Fontana}, A., {Poli}, F., {Menci}, N., {Nonino}, M., {Giallongo}, E.,
  {Cristiani}, S.,  \& {D'Odorico}, S. 2003, \apj, 587, 544

\bibitem[\protect\citeauthoryear{{Gallart}, {Zoccali}, \& {Aparicio}}{{Gallart}
  et~al.}{2005}]{gallart2005}
{Gallart}, C., {Zoccali}, M.,  \& {Aparicio}, A. 2005, \araa, 43, 387

\bibitem[\protect\citeauthoryear{{Giavalisco} et~al.}{{Giavalisco}
  et~al.}{2004}]{giavalisco2004}
{Giavalisco}, M., et~al. 2004, \apjl, 600, L103

\bibitem[\protect\citeauthoryear{{Girardi} et~al.}{{Girardi}
  et~al.}{2002}]{girardi2002}
{Girardi}, L., {Bertelli}, G., {Bressan}, A., {Chiosi}, C., {Groenewegen},
  M.~A.~T., {Marigo}, P., {Salasnich}, B.,  \& {Weiss}, A. 2002, \aap, 391, 195

\bibitem[\protect\citeauthoryear{{Girardi} et~al.}{{Girardi}
  et~al.}{2010}]{girardi2010}
{Girardi}, L., et~al. 2010, \apj, submitted

\bibitem[\protect\citeauthoryear{{Heavens} et~al.}{{Heavens}
  et~al.}{2004}]{heavens2004}
{Heavens}, A., {Panter}, B., {Jimenez}, R.,  \& {Dunlop}, J. 2004, \nat, 428,
  625

\bibitem[\protect\citeauthoryear{{Hernquist} \& {Springel}}{{Hernquist} \&
  {Springel}}{2003}]{hernquist2003}
{Hernquist}, L.,  \& {Springel}, V. 2003, \mnras, 341, 1253

\bibitem[\protect\citeauthoryear{{Holtzman} et~al.}{{Holtzman}
  et~al.}{2009}]{holtzman2010}
{Holtzman}, J., {Dalcanton}, D., J.J.~{Garnett}, {Sarajedini}, A.,  \&
  {Williams}, B.~F. 2009, \aj, submitted

\bibitem[\protect\citeauthoryear{{Hopkins}}{{Hopkins}}{2004}]{hopkins2004}
{Hopkins}, A.~M. 2004, \apj, 615, 209

\bibitem[\protect\citeauthoryear{{Hopkins}}{{Hopkins}}{2007}]{hopkins2007}
{Hopkins}, A.~M. 2007, \apj, 654, 1175

\bibitem[\protect\citeauthoryear{{Hopkins}, {Irwin}, \& {Connolly}}{{Hopkins}
  et~al.}{2001}]{hopkins2001}
{Hopkins}, A.~M., {Irwin}, M.~J.,  \& {Connolly}, A.~J. 2001, \apjl, 558, L31

\bibitem[\protect\citeauthoryear{{Ichikawa} et~al.}{{Ichikawa}
  et~al.}{1995}]{ichikawa1995}
{Ichikawa}, T., {Yanagisawa}, K., {Itoh}, N., {Tarusawa}, K., {van Driel}, W.,
  \& {Ueno}, M. 1995, \aj, 109, 2038

\bibitem[\protect\citeauthoryear{{Iwata} et~al.}{{Iwata}
  et~al.}{2003}]{iwata2003}
{Iwata}, I., {Ohta}, K., {Tamura}, N., {Ando}, M., {Wada}, S., {Watanabe}, C.,
  {Akiyama}, M.,  \& {Aoki}, K. 2003, \pasj, 55, 415

\bibitem[\protect\citeauthoryear{{Kennicutt} et~al.}{{Kennicutt}
  et~al.}{2003}]{kennicutt2003}
{Kennicutt}, R.~C., Jr., et~al. 2003, \pasp, 115, 928

\bibitem[\protect\citeauthoryear{{Kroupa}}{{Kroupa}}{2002}]{kroupa2002}
{Kroupa}, P. 2002, Science, 295, 82

\bibitem[\protect\citeauthoryear{{Lacey} et~al.}{{Lacey}
  et~al.}{2009}]{lacey2009}
{Lacey}, C.~G., {Baugh}, C.~M., {Frenk}, C.~S., {Benson}, A.~J., {Orsi}, A.,
  {Silva}, L., {Granato}, G.~L.,  \& {Bressan}, A. 2009, ArXiv e-prints

\bibitem[\protect\citeauthoryear{{Lanzetta} et~al.}{{Lanzetta}
  et~al.}{2002}]{lanzetta2002}
{Lanzetta}, K.~M., {Yahata}, N., {Pascarelle}, S., {Chen}, H.,  \&
  {Fern{\'a}ndez-Soto}, A. 2002, \apj, 570, 492

\bibitem[\protect\citeauthoryear{{Lilly}, {Carollo}, \& {Stockton}}{{Lilly}
  et~al.}{2003}]{lilly2003}
{Lilly}, S.~J., {Carollo}, C.~M.,  \& {Stockton}, A.~N. 2003, \apj, 597, 730

\bibitem[\protect\citeauthoryear{{Lilly} et~al.}{{Lilly}
  et~al.}{1996}]{lilly1996}
{Lilly}, S.~J., {Le Fevre}, O., {Hammer}, F.,  \& {Crampton}, D. 1996, \apjl,
  460, L1

\bibitem[\protect\citeauthoryear{{Madau} et~al.}{{Madau}
  et~al.}{1996}]{madau1996}
{Madau}, P., {Ferguson}, H.~C., {Dickinson}, M.~E., {Giavalisco}, M.,
  {Steidel}, C.~C.,  \& {Fruchter}, A. 1996, \mnras, 283, 1388


\bibitem[\protect\citeauthoryear{{Marigo} et~al.}{{Marigo}
  et~al.}{2008}]{marigo2008}
{Marigo}, P., {Girardi}, L., {Bressan}, A., {Groenewegen}, M.~A.~T., {Silva},
  L.,  \& {Granato}, G.~L. 2008, \aap, 482, 883

\bibitem[\protect\citeauthoryear{{Melbourne} et~al.}{{Melbourne}
  et~al.}{2010}]{melbourne2010}
{Melbourne}, J., {Williams}, B., {Dalcanton}, J., {Ammons}, S.~M., {Max}, C.,
  {Koo}, D.~C., {Girardi}, L.,  \& {Dolphin}, A. 2010, \apj, 712, 469

\bibitem[\protect\citeauthoryear{{Neistein}, {van den Bosch}, \&
  {Dekel}}{{Neistein} et~al.}{2006}]{neistein2006}
{Neistein}, E., {van den Bosch}, F.~C.,  \& {Dekel}, A. 2006, \mnras, 372, 933

\bibitem[\protect\citeauthoryear{{Puche}, {Carignan}, \& {van Gorkom}}{{Puche}
  et~al.}{1991}]{puche1991}
{Puche}, D., {Carignan}, C.,  \& {van Gorkom}, J.~H. 1991, \aj, 101, 456

\bibitem[\protect\citeauthoryear{{Reddy} et~al.}{{Reddy}
  et~al.}{2008}]{reddy2008}
{Reddy}, N.~A., {Steidel}, C.~C., {Pettini}, M., {Adelberger}, K.~L.,
  {Shapley}, A.~E., {Erb}, D.~K.,  \& {Dickinson}, M. 2008, \apjs, 175, 48

\bibitem[\protect\citeauthoryear{{Salpeter}}{{Salpeter}}{1955}]{salpeter1955}
{Salpeter}, E.~E. 1955, \apj, 121, 161

\bibitem[\protect\citeauthoryear{{Schiminovich} et~al.}{{Schiminovich}
  et~al.}{2005}]{schiminovich2005}
{Schiminovich}, D., et~al. 2005, \apjl, 619, L47

\bibitem[\protect\citeauthoryear{{Simien} \& {de Vaucouleurs}}{{Simien} \& {de
  Vaucouleurs}}{1986}]{simien1986}
{Simien}, F.,  \& {de Vaucouleurs}, G. 1986, \apj, 302, 564

\bibitem[\protect\citeauthoryear{{Springel} \& {Hernquist}}{{Springel} \&
  {Hernquist}}{2003}]{springel2003}
{Springel}, V.,  \& {Hernquist}, L. 2003, \mnras, 339, 312

\bibitem[\protect\citeauthoryear{{Steidel} et~al.}{{Steidel}
  et~al.}{1999}]{steidel1999}
{Steidel}, C.~C., {Adelberger}, K.~L., {Giavalisco}, M., {Dickinson}, M.,  \&
  {Pettini}, M. 1999, \apj, 519, 1

\bibitem[\protect\citeauthoryear{{Thomas} et~al.}{{Thomas}
  et~al.}{2005}]{thomas2005}
{Thomas}, D., {Maraston}, C., {Bender}, R.,  \& {Mendes de Oliveira}, C. 2005,
  \apj, 621, 673

\bibitem[\protect\citeauthoryear{{van den Bergh}}{{van den
  Bergh}}{2000}]{vandenbergh2000}
{van den Bergh}, S. 2000, \pasp, 112, 529

\bibitem[\protect\citeauthoryear{{Weisz} et~al}{{Weisz} et~al}{2011a}]{weisz2011a} {Weisz}, D.~R., et al. 2011a, arXiv:1101.1093

\bibitem[\protect\citeauthoryear{{Weisz} et~al}{{Weisz} et al.}{2011b}]{weisz2011} {Weisz}, D.~R., et al. 2011b, arXiv:1101.1301 

\bibitem[\protect\citeauthoryear{{Williams} et~al.}{{Williams}
  et~al.}{2009}]{williams2008}
{Williams}, B.~F., et~al. 2009, \aj, 137, 419

\bibitem[\protect\citeauthoryear{{Young} et~al.}{{Young}
  et~al.}{1996}]{young1996}
{Young}, J.~S., {Allen}, L., {Kenney}, J.~D.~P., {Lesser}, A.,  \& {Rownd}, B.
  1996, \aj, 112, 1903

\end{thebibliography}
\end{document}